\documentclass[12pt]{article}
\usepackage{latexsym,amsfonts,amsmath}
\usepackage{float}
\usepackage{multicol}
\usepackage{amssymb}
\usepackage[dvips]{graphicx,color}
\usepackage{graphicx}
\usepackage{subfigure}

\title{ Computational Aspects of Lattice QCD}
\author{Tereza Mendes\thanks{Talk presented at the
        First Brazilian Meeting in Theoretical and Computational Physics,
        Bras\'\i lia, April 6--9, 2003.}\\
        {\em \normalsize Instituto de F\'\i sica de S\~ao Carlos,
        Universidade de S\~ao Paulo}\\
        {\em \normalsize C.P. 369, CEP 13560-970, S\~ao Carlos SP}}

\begin{document}

\maketitle
\begin{abstract}
Monte Carlo simulations applied to the lattice formulation of
quantum chromodynamics (QCD) enable a study of the theory from
first principles, in a nonperturbative way.
After over two decades of developments in the methodology for this
study and with present-day computers in the teraflops range, 
lattice-QCD simulations are now able to provide quantitative predictions
with errors of a few percent. This means that these simulations 
will soon become the main source of theoretical results for comparison
with experiments in physics of the strong interactions.
It is therefore an important moment for the beginning of Brazilian
participation in the field.
\end{abstract}

\section{Introduction}
Quantum chromodynamics (QCD) is the theory describing the strong
interactions, which occur between hadrons (e.g.\ protons and neutrons)
\cite{Moriyasu}. 
The description is based on a model of elementary particles ---
the {\bf quarks} --- possessing ``color charge'' and interacting
through the exchange of gauge fields --- the {\bf gluons}.
QCD is a quantum field theory, with local $SU(3)$ gauge symmetry,
corresponding to three possible colors.
The theory is written in simple and elegant form. Its only parameters
are the masses of the various types (called ``flavors'') of
quarks considered and the value of the strong coupling constant.
Except for the symmetry under the $SU(3)$ gauge group [instead
of the $U(1)$ group], the form of the QCD Lagrangian is the same as
the one of quantum electrodynamics (QED), with the quarks corresponding
to the electrons and the gluons to the photons. (The former two are
spin-$1/2$ fermions and the latter two are massless vector bosons.)
Analogously, the strong coupling constant $\alpha_s$ corresponds
to the fine structure constant $\alpha\approx 1/137$.
The fact that the gauge group of QCD is non-Abelian introduces,
however, qualitative differences between the two theories, reflecting
the differences between the strong interactions and the electromagnetic
interactions. In particular, one obtains that the gluons possess
color charge and therefore interact with each other, as opposed to the
photons.

An important characteristic of the strong interaction is that the
coupling constant $\alpha_s$ becomes negligible only in the limit
of small distances, or equivalently in the limit of high energy or
momentum. This property is called {\bf asymptotic freedom}.
At larger distances (i.e.\ smaller energies) there is an increase
in the intensity of the interaction and it is believed that at large
distances the force of attraction between quarks is constant,
determining the {\bf confinement} of quarks and gluons inside the
hadrons.
The fact that $\alpha_s$ is not negligible at low energies makes the
study of important phenomena such as the mechanism of quark confinement,
the hadron mass spectrum and the deconfining transition at finite
temperature inaccessible to calculations using perturbation theory,
which is based on a weak-coupling expansion.
These phenomena must therefore be studied in a {\bf nonperturbative}
way.

The nonperturbative study of QCD is possible in the lattice
formulation of the theory \cite{rothe}.
In this formulation --- which consists in quantization by means
of path integrals, in the continuation to imaginary or Euclidean
time and in the lattice regularization (given by the discretization
of space-time) --- the theory becomes equivalent to a model in classical
statistical mechanics.
The continuum limit, in which physical results are obtained,
is given by the critical point of this model, which may be studied
through the usual methods of statistical mechanics.
In particular, one may perform numerical simulations by {\bf Monte Carlo}
methods, which are based on a stochastic description of the systems
considered \cite{MC}.
Due to the greater complexity of the interaction and to the larger
number of degrees of freedom, these simulations are much more elaborate
for QCD than for the usual models in statistical mechanics,
requiring considerable computational resources.
In fact, one usually needs to simulate on powerful parallel 
supercomputers, some of which were designed and built specifically
for the study of lattice QCD, such as the {\tt QCDSP} in the USA,
the {\tt Hitachi/CP-PACS} in Japan and the
{\tt APE-Mille} in Europe, all with performance in the teraflops range.
Only recently has simulation of QCD on systems of small computers,
the so-called {\bf PC clusters}, become possible.
These systems do not yet provide the same efficiency in parallelization
as the machines with parallel architecture, but their cost is much
lower.
In addition to the computational power, the numerical and analytical 
techniques used in the simulations and in the interpretation of
the produced data are of great importance in the field.
Significant progress has been achieved through the development of
more efficient simulation algorithms, new methods for interpolation and
extrapolation of the numerical data and a better understanding of
the systematic effects to which the simulation may be subject, such
as finite-volume effects and discretization errors.

There is presently great interest in the results of the simulations
described above and one hopes to be able to solve many theoretical
questions about QCD and the standard model
\cite{wilczek}.
In fact, despite the great computational difficulty, numerical studies
of QCD have provided important contributions recently, such as
accurate calculations of the strong coupling constant \cite{Hinc}
and of the hadronic mass spectrum \cite{cppacs}.
In particular, lattice simulations constitute the only known evidence
for the quark-deconfining transition at finite temperature \cite{karsch}
and its predictions are of direct interest for the current experiments
in search of new states of matter in the laboratories Brookhaven
and CERN.

Simulations of the so-called full QCD --- i.e.\ including effects
of dynamical fermions --- for quark masses in the region of physical
values are still extremely slow. They must in general be carried out on
supercomputers as the ones mentioned above and involving the effort
of large collaborations, such as the UKQCD in the United Kingdom and
the JLQCD in Japan.
The methods used in these simulations, which take on average several
months or even a few years, are often developed in studies of 
simplified versions of the theory, such as pure QCD --- the so-called
{\em quenched} approximation, in which effects of dynamical fermions 
are neglected --- and the pure $SU(2)$ theory, or models in lower
dimensions.
The consideration of this type of problems and the use of PC clusters
is the goal of our research group in the IFSC--USP.

\section{The lattice formulation}
A difficulty in the study of QCD, common to virtually all quantum field
theories, is the appearance of ultraviolet divergences (i.e.\ divergences
for high energies or short distances) in the calculation of physical
quantities \cite{LeBellac}.
Only after these ``infinities'' are removed by means of some 
renormalization procedure can one obtain finite results, which may be
compared to experiment.
It is therefore necessary to regularize the theory first, writing it 
in such a way that the singularities are isolated, and then to remove these
singularities through a redefinition of the parameters in the Lagrangian.
The lattice QCD formulation, introduced in 1974 by Wilson
\cite{wilson}, offers a convenient nonperturbative regularization,
preserving the theory's gauge invariance.
Quarks are represented at lattice sites, while gluons are represented on the 
links between neighboring sites. 
The gluonic fields are given by $SU(3)$ matrices.
The lattice action is written in terms of products of link variables along
closed loops, so that the gauge symmetry of the original action is preserved.
An excellent introduction to lattice QCD is Ref.\ \cite{rothe}.
The essential ingredients for the lattice formulation are:
\begin{enumerate}
\item Feynman's path integral formalism, in which expectation values are written
for the observables of interest as integrals over all the degrees of
freedom of the problem, with a statistical weight given by the exponential of 
the theory's classical action.
\item The Euclidean formulation, obtained by analytic continuation of the time
variable to imaginary times. In this way the (complex) oscillatory exponential
present in the integrals described above becomes real and may be interpreted
as a probability distribution.
\item The introduction of a discrete lattice for the space-time.
Correspondingly, differential operators are rewritten as finite differences 
of the discretized fields.
\end{enumerate}
The combination of the first two ingredients highlights the equivalence
between quantum field theories and classical statistical mechanics:
in Euclidean space a path integral for the quantum theory is equivalent to
a thermodynamic average for the corresponding statistical mechanical system.
For QCD, the square of the bare coupling constant $g_0$ of the field theory
corresponds directly to the temperature $1/\beta$ of the statistical
mechanical model.

The third ingredient --- the lattice discretization --- represents an
ultraviolet regularization.
In fact, the lattice spacing $a$ corresponds to a high-momentum cutoff, 
since momenta higher than $\sim 1/a$ cannot be represented on the lattice.
In this way the modes causing divergences are suppressed and the theory is
well defined. In order to recover the continuum-space theory we must take
the limit $a \to 0$.
In this process it is necessary to ``tune'' the bare parameters of the theory
--- e.g.\ the bare coupling constant $g_0$ --- in such a way that
physical quantities (correlation functions, masses, etc.) converge to
finite values, which can then be compared to experiment.
In particular, a correlation length $\xi$ (corresponding to an inverse
mass) measured in physical units --- e.g.\ fermi --- must approach
a finite limit when the lattice spacing $a$ (measured in fermi) goes
to zero. This means that the correlation length measured in units of
the lattice spacing $\xi/a$ must go to infinity.
In other words, the lattice theory considered must approach a critical
point, i.e.\ a second order phase transition.
Thus, the study of the continuum limit in quantum field theories on the
lattice is analogous to the study of critical phenomena in statistical
mechanics.
The correspondence between Euclidean field theories and classical statistical
mechanics allows the application of usual statistical-physics methods
to the study of QCD. One may use, for example, high- and low-temperature
expansions, corresponding respectively to strong- and weak-coupling
expansions for the field theory.
Another example of interplay between field theories and statistical
mechanics is the renormalization-group method, developed in parallel
for both fields \cite{LeBellac}.
A particularly important statistical technique, especially for QCD,
is Monte Carlo simulation, which allows a nonperturbative study 
of the models considered.

\section{Monte Carlo simulations}

Monte Carlo methods are generally used to sample 
the Boltzmann distribution for a statistical system 
in a stochastic way \cite{MC}.
One generates on the computer $N$ configurations for the system,
in such a way that each configuration is chosen with a probability
given by its Boltzmann weight.
The average of an observable over the $N$ configurations produced
converges in the $N\to \infty$ limit to the expectation value (or 
thermodynamic average) of this observable. For large (finite) values of
$N$ we get central values for averages of the observables of interest,
with a statistical error proportional to $1/\sqrt{N}$.
It is therefore possible to obtain an arbitrarily small error by increasing
the number of configurations produced, i.e.\ by increasing the computational
effort. The Boltzmann distribution may be defined through a model for a
physical system, especially if this system can be discretized, either
naturally or by an approximation. In many examples the art of simulation
is in the appropriate choice (or invention) of a model, for example in the
modeling of polymers by random walks.
In the case of lattice gauge theories, as explained in the previous section,
the Boltzmann distribution is obtained directly from the Lagrangian
(or equivalently the action) of the theory, without any approximations
other than the discretization of space-time, allowing therefore a
nonperturbative study from first principles.

In order to generate configurations with the desired statistical weight,
one usually introduces a Markov dynamics for the system considered, in such a way
that the resulting Markov chain has the appropriate Boltzmann distribution 
as its equilibrium probability distribution.
In this approach the thermodynamic averages described above are calculated
as time averages in the chosen dynamics. At each ``instant of time'' we generate
a new configuration of the system, in a manner that respects the appropriate
(equilibrium) distribution. Except for this restriction, the updates that determine
the dynamics may be chosen in the most convenient way, without necessarily
coinciding with the physical dynamic behavior of the system out of equilibrium.
We start from a general initial condition and follow the time sequence by
successively applying the updating procedure, which generates a new configuration 
starting from the present one.
It is common therefore to think of the simulated system as evolving by itself and
one frequently says that the observables are being ``measured'' instead of calculated.
There is also a statistical error associated with the stochastic method, as mentioned
above, and one must apply error analysis to determine the final precision of the
results. These are characteristic features of experimental studies and the methods used
for data analysis are often the same. We must remember however that we are dealing
with {\em numerical} ``experiments'', obtained from a theory (in the case of QCD) or from
a model for a physical system (in the case of statistical mechanics).

The simulation of lattice gauge theories, QCD in particular, constitutes one of the
most intensive fields of application of Monte Carlo simulations \cite[Cap.\ 11]{MC}.
As said in the Introduction, this type of study is crucial for QCD, since a perturbative
treatment of the theory is not possible for many relevant energy regions.
Despite the similarity of the methods, the Monte Carlo simulation of gauge theories
is much more complex than in the case of the usual statistical mechanical models,
requiring great computational effort and specific numerical techniques for the
production of the data.
Moreover, the physical interpretation of the generated data depends
on a correct extrapolation to the continuum limit, i.e.\ it is necessary to
``go back'' to the continuum space after the lattice simulation. More specifically,
we must consider three limits in order to obtain the desired physical results from
the simulation data:

\begin{itemize}
\item {\bf The infinite-volume limit (or thermodynamic limit):}
Just as in statistical mechanics, simulations of QCD are carried out for
finite lattice volumes, since the computer's memory is finite. The lattice
volume must therefore be sufficiently large with respect to the
physical distance that is relevant to the problem at hand, so that
finite-volume effects are not significant.
(Correspondingly, one may not consider energies or momenta that are too low,
since the finite lattice is equivalent to an infrared cutoff.)
Finite-volume effects can in general be estimated through a 
{\em finite-size-scaling} analysis of the data.
For QCD simulations it generally would suffice to consider lattice
sizes $L\approx 7\,fm$.
\item {\bf The continuum limit:} In order to recover the original 
continuum physics it is necessary to take the limit $a\to 0$.
At the same time, the calculated quantities must be renormalized, i.e.\ 
redefined so as to generate finite results in the continuum limit.
This can be done nonperturbatively by using the physical values of some
observables, which are known experimentally.
For example, writing the pion mass calculated on the lattice as
$m_{\pi}\,a$ (where $m_{\pi}$ is the physical mass), we obtain the value
of the ultraviolet cutoff $a$ in physical units. The other calculated
quantities may then be written in the same form in terms of $a$
(which tends to zero) and ``translated'' into physical units, generating
(physical) finite values.
In practice, the value of $a$ must be sufficiently small when compared to the
relevant distance for the problem, for example $a\approx 0.05\,fm$.
It is important to notice that several discretizations of the action are
possible and that one may consider the so-called {\em improved actions},
which converge to the continuum limit faster, i.e.\ for larger values
of $a$. (One may also consider several different discretizations for the
fermionic fields.)
\item {\bf The chiral limit:} It is very hard to consider physical values 
for the quark masses in the numerical simulations.
That happens especially for the light quarks (up and down), whose masses
are close to zero, the so-called chiral limit.
The simulations are usually done for larger masses and the results are then
extrapolated using chiral perturbation theory.
\end{itemize}

The above limits are not independent, since to get to the continuum limit
and to be able to consider small masses for the quarks one needs a 
sufficiently large number of lattice points (corresponding to a small enough
lattice spacing and to a large enough physical size of the lattice),
which increases considerably the computational effort.
For example, with the present algorithms, a numerical simulation at the
ideal values of $L$ and $a$ given above and for physical values of the quark
masses would take approximately \cite{sharpe} 300 years on a super-computer
with $1 T\!flops$ power, corresponding to the fastest computers available 
today.\footnote{One teraflops ($T\!flops$) is equivalent to $10^{12}$
floating-point operations per second.}
Employing improved actions and making use of the chiral-limit extrapolation,
the same simulation can be done in 2 years.
These simulations are much faster for the so-called {\em quenched} case, in which
the configurations are produced considering quarks of infinite mass, i.e.\ 
without considering effects of dynamical quarks.
(Note that the observables calculated for each configuration may still include
quarks with the desired masses, which are then called {\em valence quarks}.)
Despite being a rough (and uncontrolled) approximation, one verifies that in
many cases the quenched approximation shows small corrections with respect to
the full QCD, indicating that in these cases the effect of dynamical fermions
is small.
Today, quenched simulations can be done with good precision and there
are also several ongoing simulations of QCD with dynamical quarks, for
several discretizations of the Dirac operator.

\section{Progresses in lattice QCD}

Progresses in the field are reported annually at the {\em Lattice} 
conference \cite{lattice}. We describe briefly below three important
contributions of lattice QCD to the confirmation/prediction of 
experimental results.

\begin{itemize}
\item {\bf Strong coupling constant:}
The strong coupling constant $\alpha_s(\mu_0)$, taken at a fixed
reference scale $\mu_0$, is the only free parameter of QCD and must therefore
be known to the highest precision possible. Numerical simulations of QCD
are now able to produce calculations of $\alpha_s$ with precision comparable
to the experimental one or better. These results are presently included
in the world average for this quantity \cite{Hinc}.
The value of $\alpha_s$ should also be determined with good precision over
as large a range of values as possible, describing the behavior of the 
interaction from the nonperturbative (strong-coupling) to the perturbative 
(weak-coupling) regime. Several methods for the calculation of the {\em running} 
strong coupling are being pursued, both in the quenched approximation and in 
the full-QCD case. A recent review of these methods can be found in 
\cite{attilio}.

\item {\bf Hadron spectrum:}
It is possible to obtain physical values of hadron masses from lattice QCD
simulations, as described in the previous section.
To this end one must tune the $n_f+1$ parameters of the theory, where $n_f$ 
is the number of quark flavors considered.
Thus, one must use $n_f+1$ known experimental results as inputs and the 
subsequent calculations are physical predictions of the numerical simulation.
The mass spectrum of the light hadrons (including the two light quarks and the
strange quark) has been determined for the quenched case with great 
precision in \cite{cppacs}.
One does not obtain complete agreement with the experimental spectrum, but the
observed discrepancies are of at most 10\%. We conclude then that the difference
introduced by the quenched approximation is only quantitative.
Similar calculations are now being performed for the full-QCD case.

\item {\bf QCD phase transition:}
The predicted phase transition for QCD at high temperatures is clearly
observed in lattice QCD simulations \cite{karsch}. For the quenched case 
one studies the deconfining transition itself, while for the full-QCD case
one must consider the transition associated with the restoration of the
chiral symmetry. This is an exact symmetry of the QCD Lagrangian in the limit of
zero masses for the quarks and is spontaneously broken at low temperatures.
Studies of pure QCD (i.e.\ the quenched approximation) are done with high
precision for the determination of the critical temperature and of the 
thermodynamic equation of state for the system. In this case one encounters
qualitative differences between pure QCD and the full-QCD case.
(In particular, the order of the phase transition is different for the
two cases.)

\end{itemize}

\section{Lattice QCD at the IFSC--USP}

Since the beginning of 2001 we have been carrying out a project
on numerical simulations of lattice gauge theories at the 
Physics Department of the University of S\~ao Paulo in S\~ao Carlos 
(IFSC--USP), funded by FAPESP \cite{noi}.
The project included the installation of a PC cluster with 20
processing nodes (16 nodes with 866 MHz Pentium III CPU,
4 nodes with 1.7 GHz Pentium 4 CPU, all with 256 MB of 
memory).\footnote{The resulting computer power is of approximately
$20\,G\!flops$ for peak performance.}
We have performed production runs since July of 2001 and have started
intensive parallel simulations in 2003.
We consider applications that require moderate computer power,
such as several problems for the $SU(2)$ gauge theory.
In particular, we propose a new method for the study of the running 
coupling constant $\alpha_s$,
based on the calculation of gluon and ghost propagators 
\cite{attilio,Bloch:2002we}.
We also carry out numerical studies of the infrared properties of QCD
(through the study of the behavior of gluon and ghost propagators in
the infrared limit) \cite{Cucchieri:2003di}, of gauge-fixing techniques
\cite{Cucchieri:2003fb},
of the chiral phase transition of QCD with two dynamical fermions 
\cite{mendes_bento}
[for which one expects analogies with the $O(4)$ spin model],
of the electroweak phase transition and aspects of spin models and 
percolation.

\section*{Acknowledgements}
The author thanks Attilio Cucchieri for the help in the preparation
of this manuscript.
Research supported by FAPESP, Proc.\ No.\ 00/05047-5.

{}


\begin{thebibliography}{}

\bibitem{Moriyasu} {\em An Elementary Primer for Gauge Theory},
       K. Moriyasu, (World Scientific, Singapore, 1983).

\bibitem{rothe} {\em Lattice gauge theories. An introduction},
                  H.J. Rothe, (World Scientific, Singapore, 1997).

\bibitem{MC} {\em A guide to Monte Carlo simulations in Statistical
              Physics}, D.P. Landau and K. Binder,
              (Cambridge University Press, Cambridge, 2000).

\bibitem{wilczek} {\em Opportunities, challenges, and phantasies 
       in lattice QCD}, F. Wilczek, presented at {\em Lattice 2002},
       Boston, USA, June 24--29, 2002,
       {\tt http://} {\tt arXiv.org/abs/hep-lat/0212041}.

\bibitem{Hinc} {\em Quantum chromodynamics}, I.\ Hinchliffe, in
               {\em Review of particle physics}, Particle Data Group
               (K. Hagiwara et al.), Phys.\ Rev.\ D 66 (2002) 010001.

\bibitem{cppacs} {\em Quenched light hadron spectrum},
       S. Aoki et al.\ (CP-PACS collaboration),
       Phys.\ Rev.\ Lett.\ 84, 238 (2000).

\bibitem{karsch} {\em Lattice QCD at high temperature and density},
       F.~Karsch, Lect.\ Notes Phys.\ 583, 209 (2002).

\bibitem{LeBellac} {\em Quantum and Statistical Field Theory},
                   M. Le Bellac, (Oxford University Press,
                   Oxford, 1995).

\bibitem{wilson} {\em Confinement of Quarks},
              K.G.~Wilson, Phys.\ Rev.\ D10, 2445 (1974).

\bibitem{lattice} Proceedings of the {\em International Symposium
                  on lattice field theory}, Nucl.\ Phys.\ B
                  (Proc.\ Suppl.) 106--107 (2002) and previous years.

\bibitem{attilio} {\em Lattice simulations for the running 
       coupling constant of QCD}, 
       A. Cucchieri, presented at {\em Hadron Physics 2002},
       Bento Gon\c calves, April 14--19, 2002,
       {\tt http://arXiv.org/abs/hep-lat/0209076}.

\bibitem{sharpe} {\em Progress in lattice gauge theory},
       S.R. Sharpe, in {\em Vancouver 1998, High energy physics}, 
       vol.\ 1, 171--190,
       {\tt http://arXiv.org/abs/hep-lat/} {\tt 9811006}. 

\bibitem{noi} {\tt http://lattice.if.sc.usp.br/}.

\bibitem{Bloch:2002we}
{\em Running coupling constant and propagators in SU(2) Landau gauge},
J.C. Bloch, A. Cucchieri, K. Langfeld and T. Mendes,
Nucl.\ Phys.\ B  (Proc.\ Suppl.) 119, 736 (2003), 
{\tt http://arXiv.org/abs/hep-lat/0209040}.

\bibitem{Cucchieri:2003di}
{\em $SU(2)$ Landau gluon propagator on a $140^3$ lattice},
A.~Cucchieri, T. Mendes and A.R. Taurines,
Phys.\ Rev.\ D {\bf 67}, 091502 (2003)

\bibitem{Cucchieri:2003fb}
{\em Critical slowing-down in $SU(2)$ Landau-gauge-fixing algorithms at 
$\beta = \infty$},
A.~Cucchieri and T.~Mendes,
Comput.\ Phys.\ Commun.\  {\bf 154}, 1 (2003).

\bibitem{mendes_bento}
{\em Study of scaling at the chiral phase transition using
    2-flavor lattice QCD data}, T. Mendes,
    to appear in the Proceedings of {\em Hadron Physics 2002},
    Bento Gon\c calves, April 14--19, 2002.

\end{thebibliography}
\end{document}